\documentclass[floatfix, aip, rsi, amsmath,amssymb, preprint, reprint, usenames,dvipsnames ]{revtex4-1}

\usepackage{graphicx}
\usepackage{dcolumn}
\usepackage{bm}
\usepackage{gensymb}
\usepackage{soul,xcolor}

\usepackage{ulem}

\setstcolor{NavyBlue}

\begin{document}

\title[]{Unravelling the Origins of Ice Nucleation on Organic Crystals}

\author{Gabriele C. Sosso}
\affiliation{Department of Chemistry and Centre for Scientific Computing, University of Warwick, Gibbet Hill Road, Coventry CV4 7AL, United Kingdom}
\email{G.Sosso@warwick.ac.uk}
\author{Thomas F. Whale}
\affiliation{Institute for Climate and Atmospheric Science, School of Earth and Environment, University of Leeds, Leeds LS2 9JT, United Kingdom}
\author{Mark A. Holden}
\affiliation{School of Earth and Environment and School of Chemistry, University of Leeds, Leeds LS2 9JT, United Kingdom}
\author{Philipp Pedevilla}
\affiliation{Thomas Young Centre, London Centre for Nanotechnology
and Department of Physics and Astronomy, University College London, Gower Street London, London, WC1E 6BT, UK}
\author{Benjamin J. Murray}
\affiliation{Institute for Climate and Atmospheric Science, School of Earth and Environment, University of Leeds, Leeds LS2 9JT, United Kingdom}
\author{Angelos Michaelides}
\affiliation{Thomas Young Centre, London Centre for Nanotechnology
and Department of Physics and Astronomy, University College London, Gower Street London, London, WC1E 6BT, UK}

\begin{abstract}
Organic molecules such as steroids or amino acids form crystals
that can facilitate the formation of ice -- arguably the most important phase transition on earth.
However, the origin of the ice nucleating ability of organic crystals is still largely unknown. Here, we combine
experiments and simulations to unravel the microscopic details of ice formation on cholesterol, a prototypical organic
crystal widely used in cryopreservation. We find that cholesterol -- which is also a
substantial component of cell membranes -- is an ice nucleating agent more potent than many inorganic substrates,
including the mineral feldspar (one of the most active ice nucleating material in the atmosphere). Scanning electron
microscopy measurements reveal a variety of morphological features on the surfaces of cholesterol crystals: this
suggests that the topography of the surface is key to the broad range of ice nucleating activity observed (from -4 to -20 C\degree).
In addition, we show via molecular simulations that cholesterol crystals aid the formation of ice nuclei in a
unconventional fashion. Rather than providing a template for a flat ice-like contact layer (as found in the case of many
inorganic substrates), the flexibility of the cholesterol surface and its low density of hydrophilic functional groups
leads to the formation of molecular cages involving both water molecules and terminal hydroxyl groups of the cholesterol
surface. These cages are made of 6- and, surprisingly, 5-membered hydrogen bonded rings of water and hydroxyl groups
that favour the nucleation of hexagonal as well as cubic ice (a rare occurrence). We
argue that the phenomenal ice nucleating activity of steroids such as cholesterol (and potentially of many other
organic crystals) is due to (i) the ability of flexible hydrophilic surfaces to form unconventional ice-templating
structures and (ii) the different nucleation sites offered by the diverse topography of the crystalline
surfaces. These findings clarify how exactly organic crystals promote the formation of ice, thus paving the way toward
the understanding of ice formation in soft and biological matter - with obvious reverberations on atmospheric science
and cryobiology.

\end{abstract}

\maketitle

\section{Introduction}

The freezing of liquid water into crystalline ice is a ubiquitous phenomenon which is part of our everyday
experience~\cite{bartels-rausch_chemistry:_2013} and has countless reverberations in fields as diverse as
cryobiology~\cite{moce_use_2010,massie_gmp_2014,john_morris_controlled_2013} and atmospheric
science~\cite{murray_ice_2012,slater_blue-sky_2015}.  Strikingly, the overwhelming majority of ice on earth forms
heterogeneously, i.e. thanks to the presence of substances, other than water itself, which facilitate the ice nucleation
process~\cite{murray_ice_2012,sosso_crystal_2016}. Much of what is known about heterogeneous ice nucleation has come
from the study of atmospherically relevant ice nucleating agents~\cite{murray_ice_2012}: in fact, heterogeneous ice
nucleation from supercooled water plays a critical role in the glaciation of mixed phase clouds, which in turn
influences the climate ~\cite{lohmann_global_2005,tan_observational_2016}. A variety of substances are known to nucleate
ice efficiently in the atmosphere, including inorganic species such as silver iodide~\cite{marcolli_ice_2016},
feldspar~\cite{atkinson_importance_2013,harrison_not_2016,kiselev_active_2017} as well as biological entities such as
the bacterium {\it pseudomonas syringae}~\cite{maki_ice_1974,lindemann_ice-nucleating_1986,pandey_ice-nucleating_2016}
or birch pollen~\cite{pummer_suspendable_2012}.

Biological ice nucleating agents also play a key role in the ever-growing field of cryobiology: in fact, the formation
of ice in biological matter is the cornerstone of cryotherapy and
cryopreservation~\cite{john_morris_controlled_2013,bar_dolev_ice-binding_2016}, i.e. the long-term storage of frozen
biological material which is essential to enable cutting-edge technologies such as regenerative medicine
~\cite{asghar_preserving_2014,aijaz_biomanufacturing_2018}.  A number of organic crystals have been known to facilitate
ice nucleation~\cite{fukuta_epitaxial_1963,fukuta_experimental_1966}, and molecular crystals of steroids such as
cholesterol (CHL)~\cite{head_steroids_1961} are used to boost the formation of ice when cryopreserving biological
material~\cite{moce_use_2010,massie_cryopreservation_2011}.  Importantly, CHL molecules also represent a major component
(up to 40\%) of animal cell membranes~\cite{krause_structural_2014}, thus prompting the question of whether this steroid
can play a role as ice nucleator in the context of ice formation in biological matter.

However, the the microscopic details of heterogeneous ice nucleation on CHL - and indeed on the vast majority of organic
and inorganic compounds alike - remain remarkably poorly understood~\cite{sosso_crystal_2016}, although a substantial
body of experimental work has been devoted to assess the ice nucleation ability of biological
matter~\cite{fukuta_experimental_1966,edwards_mechanism_1971,pratt_situ_2009,hoose_how_2010,wang_deposition_2012,davies_ice-binding_2014,liu_janus_2016,campbell_observing_2017}.
In fact, the reason why many biological ice nucleating agents display a far stronger ice nucleating activity than most
inorganic materials~\cite{murray_ice_2012,sosso_crystal_2016} is still largely unknown.  Partly this is because
obtaining molecular-level insight into the nucleation process is still a formidable challenge for experiments, and only
very recently simulations of heterogeneous ice nucleation have become
feasible~\cite{sosso_crystal_2016,zielke_simulations_2016,zhang_impact_2014,bi_heterogeneous_2016,reinhardt_effects_2014,fraux_note:_2014,pedevilla_what_2017,kiselev_active_2017,lupi_reaction_2017,bi_enhanced_2017},
largely thanks to the capabilities of the coarse grained mW water model~\cite{molinero_water_2009}. Indeed mW has played
a pivotal role in enabling systematic investigations of ice nucleation on e.g.
carbonaceous~\cite{lupi_heterogeneous_2014} or hydroxylated organic surfaces~\cite{qiu_ice_2017}.  However, fully
atomistic water models and enhanced sampling methods are often required to take into account the subtleties of the
hydrogen bond network between water and complex impurities~\cite{sosso_microscopic_2016,sosso_ice_2016}.

In this work, we bring together experiments and simulations to take an ambitious step forward in furthering our
understanding of ice formation on organic crystals. We focus on CHL, due to its relevance in cryopreservation and its
role within cellular membrane, unravelling microscopic motivations for heterogeneous ice nucleation likely to be shared
by many other organic crystals.  We find via micro-litre droplet nucleation measurements ($\mu$l-NIPI) that CHL crystals
display an outstanding ice nucleation ability (stronger than most inorganic ice nucleating agents), with freezing events
initiating at very mild supercooling $\Delta T_S = T_{\text{Melt}}-T$ = 4 K down to $\Delta T_S$ = 20 K. Scanning
electron microscopy suggests that the activity of these crystals as ice nucleating agents across such a wide temperature
range could be due to the diverse topography of the surface of the cholesterol crystals, which are likely to offer a
variety of different nucleation sites. In order to get a molecular-level insight into the mechanism of ice formation on
CHL crystals, we harness enhanced sampling simulations, focusing on the hydroxylated (001) face of cholesterol
monohydrate (CHLM) -- the most relevant surface (and polymorph, as discussed below) in biological scenarios.  We find
that CHL crystals facilitate the formation of ice in a non-conventional fashion: in contrast to what has been observed
in the case of inorganic substrates such as e.g. carbonaceous particles~\cite{lupi_heterogeneous_2014} or clay
minerals~\cite{sosso_ice_2016}, the flexibility of the CHLM surface and the large spacing of its hydroxyl groups prevent
the formation of a flat, ice-like layer of water molecules at the water-crystal interface. Instead, the hydroxyl groups
participate in the formation of 5- and 6-membered hydrogen bonded rings of water molecules forming peculiar molecular
``cages'' that provide an effective template for the nucleation of {\it both} cubic and hexagonal ice (a rare
occurrence).

As a whole, our findings suggest that the formation of ice on CHL crystals originates from the ability of their flexible
hydrophilic surfaces to trigger the formation of unconventional ice-templating molecular features.  In addition,
different nucleation sites potentially offered by the diverse topography of the crystal can further enhance the
intrinsic ice nucleation potential of CHL surfaces. This insight could help to understand ice formation on a number of
other organic compounds, from amino acid crystals~\cite{evans_ice_1966,gavish_role_1992} to bacterial
fragments~\cite{wolber_identification_1986,gurian-sherman_bacterial_1993} -- as they are all characterised by the
presence of flexible hydrophilic surfaces displaying diverse topological features. In addition, organic crystals such as
cholesterol are positioned ``in between'' inorganic and biological ice nucleating agents: they possess the order and the
crystalline surfaces of the former, and the complexity and flexibility of the latter. This work thus paves the way to a
molecular-level understanding of ice formation in biological matter, tackling a substrate (CHL crystals) that embeds
unique features of very different classes of materials.

\section{Methods}

\subsection*{$\mu$l-NIPI experiments and scanning electron microscopy measurements}

\noindent The ice nucleation efficiency of CHLM was evaluated using an adapted version of the $\mu$L Nucleation by
Immersed Particles Instrument ($\mu$l-NIPI) described in detail in Ref.~\cite{whale_technique_2015}. To make the flat
plates we dissolved 2g of pure CHL (Sigma Aldrich) in approximately 30 ml of hot ($\sim$ 343 K) 95\% ethanol
(Sigma-Aldrich). The CHL solution was then allowed to cool slowly, causing crystallisation of large (up to approx. 1 cm
across) flat plates of CHLM. Individual plates of CHLM of around 2mm diameter were then recovered by vacuum filtration
on a filter membrane and placed onto a thin ($\sim$ 0.1 mm) glass plate. The glass plate was then placed onto an EF600
Stirling cryocooler. A Picus Biohit electronic pipette was then used to deposit 1 $\mu$l droplets of MilliQ water onto
the separated CHLM plates. The EF600 cryocooler was then used to reduce the temperature of the droplets at a rate of 1
K/min. freezing was monitored using a camera. In this way the droplet fraction frozen curve presented in
Fig.~\ref{FIG_1}a was built up. The data is the result of several cooling runs as only about 10 droplets could be frozen
per experiment.  It was important that plates were not in contact as ice clearly propagated across the CHL surface,
triggering neighbouring droplets after an initial freezing events, when multiple droplets were placed on a single plate.

In order to calculate the (surface) density of the active ice nucleation sites ($n_s$, commonly used to compare the
ice nucleating efficiency of different substances~\cite{murray_ice_2012}) on the CHLM surface (reported in
Fig.~\ref{FIG_1}) we have used the following expression:

\begin{equation}
        \frac{n(T)}{N} = 1 - \exp{[ - n_{s}\left( T \right)A]},
\end{equation}

\noindent where $n(T)$ is the number of droplets frozen at temperature $T$, $N$ is the total number of droplets in
the experiment and $A$ is the surface area of nucleating agent per droplet. The value of $A$ for each droplet was
measured using the image analysis software imageJ, customarily used to quantify particles size in
biosciences~\cite{schneider_nih_2012}. The resulting estimate of the mean value of $A$ is 1.82 $\pm$ 0.46 mm$^2$. 

The uncertainties associated with the values of $n_s$ have been calculated using Monte-Carlo simulations of possible
active ice nucleation sites distributions, propagated with the uncertainty associated with $A$ -- as described in
Harrison et al.~\cite{harrison_not_2016}.  These simulations generate a list of possible values for the number of active
sites per droplet for a given experiment, given the observed freezing data. By repeating this process a great many
times, a distribution of the possible active site distributions that can account for the freezing of each droplet is obtained.
The error bars for the CHLM $n_s$ data reported in Fig.~\ref{FIG_1} are generated by propagating this distribution with
the uncertainty in surface area of cholesterol per droplet and taking the 95\% confidence interval of the resulting
distribution. At high and low temperature ends of the reported data, where the Poisson uncertainty (i.e. the error originating from the
Monte-Carlo simulations) is largest, the contribution of the uncertainty in surface area amounts to approximately 25\%
of total uncertainty in $n_s(T)$, with the Poisson uncertainty in the active site distribution accounting for the remainder
of the error bars. 

Scanning electron microscopy (SEM) was performed on CHLM plates. These were mounted on copper tape, then coated with 2
nm of iridium. SEM was performed with an FEI Nova NanoSEM 450 in high vacuum mode, using an Everhart-Thornley Detector
(ETD).

\subsection*{Molecular dynamics simulations}

\noindent The computational setup we have used is depicted in Fig.~\ref{FIG_3}a. A single layer of CHL molecules,
cleaved along the (001) plane (perpendicular to the normal to the slab) was prepared by starting from the experimental
cell parameters and lattice positions~\cite{craven_crystal_1976}. Specifically, a CHLM crystal system made of two
mirroring slabs (intercalated by water molecules, in a ratio of 1:1) was cleaved along the (001) plane. The triclinc
symmetry of the system (space group $C1$) was preserved, and we have constructed a 3 by 3 supercell with in-plane
dimensions of 37.17 and 36.57 \AA.  We positioned 1923 water molecules randomly atop this CHLM slab at the density of
the TIP4P/Ice model~\cite{abascal_potential_2005} at 300 K, and expanded the dimension of the simulation cell along the
normal to the slab to 100 \AA. This setup allows for a physically meaningful equilibration of the water at the density
of interest at a given temperature, but suffers from two distinct drawbacks: i) the CHLM slab possesses a net dipole
moment which is not compensated throughout the simulation cell and ii) the presence of the water-vacuum interface can
alter the structure and the dynamics of the liquid film. However, we have previously addressed these issues in previous
work dealing with the hydroxylated (001) polar surface of the clay mineral
kaolinite~\cite{sosso_microscopic_2016,sosso_ice_2016}, concluding that such details do not affect the mechanism nor the
kinetics of ice formation. In addition, the water film is thick enough to allow a bulk-like region to exist in terms of
both structure and dynamics. The effect of the water-vacuum interface is therefore negligible. The slab considered in
this work presents the hydrophilic, -OH terminated heads of the CHL molecules to the water, in agreement with
experimental insight~\cite{abendan_surface_2002}.  As we discuss in the main text, the interaction between the hydroxyl
groups (which display amphoteric characteristics in terms of the hydrogen bond network) of CHL molecules and water is
responsible for the templating effect of CHLM crystals which serves to promote ice nucleation.

The CHARMM36~\cite{bjelkmar_implementation_2010} force field was used to model the CHL crystals, taking advantage of a
recent update of this force field parameters explicitly with respect to CHL~\cite{lim_update_2012}.  In order to mimic
the experimental conditions, we have constrained the system at the experimental lateral dimensions (detailed together
with the computational geometry in the SI), and we have also restrained the positions of the hydrophobic tail of each
CHL molecule (specifically, the carbon atoms C25, C26 and C27, see the inset of Fig. S1 in the SI) by means of an
harmonic potential characterised by a spring constant of 10000 kJ/mol. All the other atoms within the CHLM slab are
unconstrained. We have verified that the thermal expansion of the crystal at 230 K ($\sim$ 0.1\% with respect to each
lateral dimension) does not alter the structure nor the dynamics of the water-kaolinite interface. This setup is thus as
close as we can get to the realistic (001) hydrophilic surface of CHLM within the CHARMM36 model. Implications of the
flexibility of the CHLM slab are discussed in the SI.  The interaction between the water molecules have been modelled
using the TIP4P/Ice model~\cite{abascal_potential_2005}, so that our results are consistent with the homogeneous
simulations of Ref.~\citenum{haji-akbari_direct_2015}. The interaction parameters between the clay and the water were
obtained using the standard Lorentz-Berthelot mixing rules~\cite{lorentz_ueber_1881}.

Extreme care must be taken in order to correctly reproduce the structure and the dynamics of the water-CHLM interface.
The Forward Flux Sampling (FFS) simulations reported in this work rely on a massive collection of unbiased Molecular
Dynamics (MD) runs, all of which have been performed using the GROMACS package, version 4.6.7. The code was compiled in
single-precision, in order to alleviate the huge computational workload needed to converge the FFS algorithm and because
we have taken advantage of GPU acceleration, which is not available in the double-precision version. The equations of
motions were integrated using a leap-frog integrator with a time step of 2 fs. The van der Waals (non bonded)
interactions were considered up to 10 \AA, where a switching function was used to bring them to zero at 12 \AA.
Electrostatic interactions have been dealt with by means of an Ewald summation up to 12 \AA. The NVT ensemble was
sampled at 230 K using a stochastic velocity rescaling thermostat~\cite{bussi_canonical_2007} with a very weak coupling
constant of 4 ps in order to avoid temperature gradients throughout the system. The geometry of the water molecules
(TIP4P/Ice being a rigid model) was constrained using the SETTLE algorithm~\cite{miyamoto_settle:_1992} while the
P\_LINCS algorithm~\cite{hess_lincs:_1997} was used to constrain the O-H bonds within the clay.  We have verified that
these settings reproduce the dynamical properties of water reported in Ref.~\cite{haji-akbari_direct_2015}. The system
was equilibrated at 300 K for 10 ns, before being quenched to 230 K over 50 ns. This is the starting point for the
calculation of the initial flux rate for the FFS algorithm, which lasted about 1.5 $\mu$s and thus allowed us to
investigate the water-CHLM interface as well (see e.g. Fig.~\ref{FIG_3}).

\section{\label{res}Results}

\subsection{\label{r_exp} Cholesterol promotes the formation of ice across a wide range of temperatures}

\begin{figure}[t!]
        \begin{centering}
                \centerline{\includegraphics[width=8.3cm]{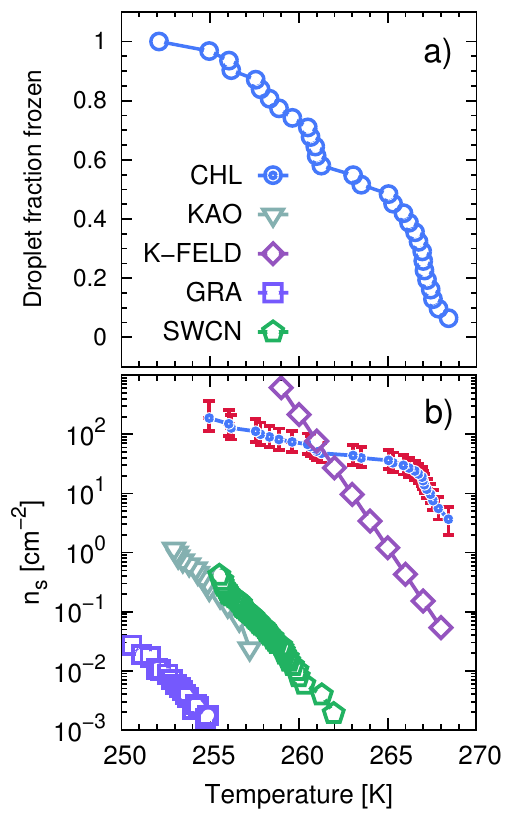}}
        \par\end{centering}
        \protect
	\caption{\textit{CHLM crystals promote the formation of ice across a wide range of temperatures.} (a) Droplet
fraction frozen as a function of temperature for 1$\mu$l water droplets placed onto a CHLM substrate. (b) Ice-active
surface site density $n_s$ (see text and Methods section) values for the same data reported for CHLM (CHL) in panel (a),
together with $n_s$ values for kaolinite (KAO) from Herbert {\it et al.}~\cite{herbert_representing_2014}, BCS376
feldspar (K-FELD) from Atkinson {\it et al.}~\cite{atkinson_importance_2013},
graphene oxide (GRA) from Whale {\it et al.}~\cite{whale_ice_2015} and
single-walled carbon nanotubes (SWCN) also from Ref.~\citenum{whale_ice_2015}. The uncertainty in terms of temperature
associated with the CHLM data is $\pm$ 0.4 K.}
        \label{FIG_1}
\end{figure}

\noindent We start by experimentally investigating the ice nucleating ability of cholesterol crystals - as a function of
supercooling. CHL can crystallise into two different polymorphs, namely anhydrous~\cite{shieh_crystal_1977} (CHLA) and
CHLM~\cite{craven_crystal_1976}. The latter is the most relevant to ice formation, as it spontaneously forms in aqueous
environments~\cite{craven_crystal_1976,loomis_phase_1979,garti_correlation_1981,rapaport_cholesterol_2001,abendan_surface_2002,solomonov_trapping_2005}.
Conveniently, CHLM crystallises from a mixture of 95\% ethanol and 5\% water as plates with the (001) surface forming
the flat surface of the plates~\cite{garti_correlation_1981}. The platey crystal habit of CHLM is characteristic of CHLM
as opposed to CHLA, which tends to crystallise as needles.  CHLM crystals display a layered structure: bilayers of CHL
molecules are stacked along the [001] direction, and facile cleavage along the (001) planes leads to surfaces exposing
either a -CH$_3$ terminated, hydrophobic surface or a -OH terminated, hydrophilic surface. Atomic and chemical force
microscopy measurements indicate that in aqueous and organic solution conditions, the hydrophilic (001) surface is most
abundantly found, in the form of largely homogeneous crystalline faces~\cite{abendan_surface_2002}. Early experimental
evidence suggested substantial ice nucleation activity of CHLM at very mild supercooling ($\Delta T_S$ = 5
K)~\cite{head_steroids_1961,head_ice_1962,fukuta_epitaxial_1963}.  The ice nucleation efficiency of CHLM was evaluated
using an adapted version of the $\mu$l-Nucleation by Immersed Particles Instrument ($\mu$l-NIPI) experiments $\mu$l-NIPI
described in detail in Ref.~\citenum{whale_technique_2015}.  Experiments were performed by placing droplets directly
onto a surface of crystalline CHLM.  We used an electronic pipette to place 1 $\mu$l droplets of water onto the (001)
plane of plates of CHLM. The water droplets were then cooled down at a rate of 1 K/min and freezing monitored using a
camera. In this way the fraction of frozen droplets can be determined as a function of temperature. Note that as the
crystalline surface is submerged in liquid water these experiments are conducted at 100\% relative humidity - i.e. in
''immersion mode``~\cite{vali_technical_2015}.

In Fig.~\ref{FIG_1}a we report the fraction of frozen droplets as a function of temperature for CHLM. It can be seen
that CHLM can induce ice nucleation at temperatures as warm as 269 K. This agrees with previous studies which have
reported high nucleation temperatures for CHL in the immersion mode~\cite{moce_use_2010,massie_cryopreservation_2011}.
In here, we investigate the ice nucleating ability of CHLM as a function of supercooling.  As shown in
Fig.~\ref{FIG_1}a, the spread of nucleation temperatures for the CHLM sheets is very broad, with some of them freezing
at temperatures as low as 252 K. To allow for a comparison of the efficiency of ice nucleation by CHLM with other known
nucleating species we have calculated the ice-active surface site density ($n_s$) for CHLM on the basis of the size of
the contact patch of the water droplets with the CHL plates. As explained in greater detail in the Methods section,
$n_s$ is a site specific measure of ice nucleation efficiency which does not take into account the time dependence of
ice nucleation, on the basis that the impact of time dependence on heterogeneous ice nucleation is generally
minimal~\cite{vali_interpretation_2014,herbert_representing_2014,vali_technical_2015}. We have compared the ice
nucleating efficiency of CHLM with that of e.g. kaolinite powder~\cite{herbert_representing_2014}, which has commonly
been regarded as an efficient ice nucleating agent in the past~\cite{pruppacher_microphysics_1997} and of BCS 376
feldspar powder, which is known to nucleate ice highly efficiently~\cite{atkinson_importance_2013} and was likely
responsible for earlier observations of efficient ice nucleation in kaolinite samples. It can be seen that CHLM
nucleates ice far more efficiently than kaolinite and more efficiently even than the feldspar at warm temperatures.
Thus, CHLM has the potential to be a highly efficient ice nucleating agent in immersion mode across a wide range of
temperatures - which is the scenario typically encountered when dealing with cryobiological applications.

\subsection{\label{r_exp2} The role of surface topography}

\begin{figure}[t!]
        \begin{centering}
                \centerline{\includegraphics[width=8.3cm]{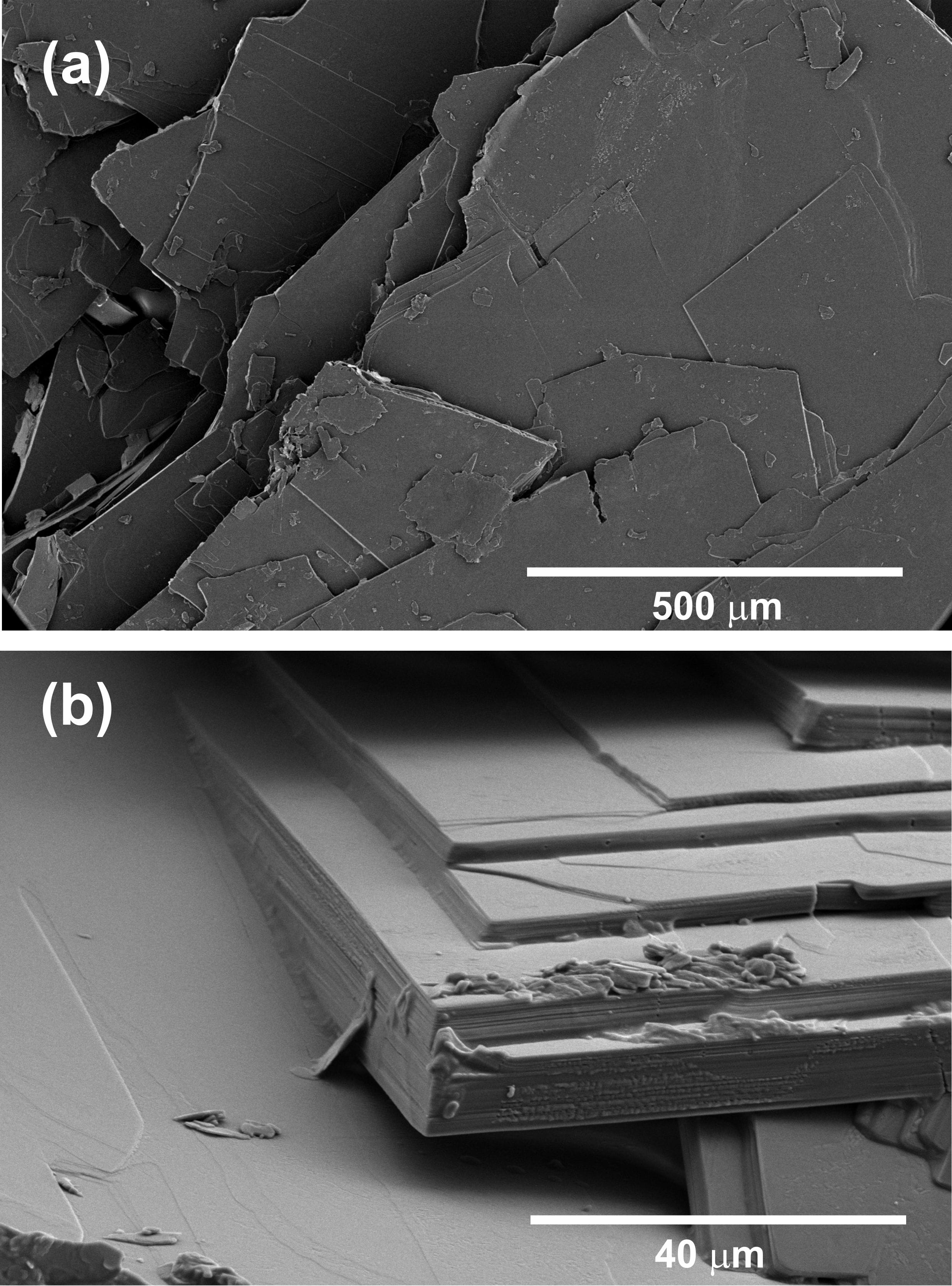}}
        \par\end{centering}
        \protect
        \caption{\textit{CHLM crystals display a diverse surface topography.} SEM images of CHLM crystals, which
predominantly expose \{001\} surfaces -- consistent with what has been reported in Ref.~\citenum{abendan_surface_2002}.
It is quite clear that these plates possess numerous topographical features, particularly steps and terraces.}
\label{FIG_2} \end{figure}

As noted above the spread of freezing temperatures we report for CHLM is very broad. This behaviour suggests that the
nucleation behaviour of CHLM is spatially heterogeneous, i.e. different parts of the surface nucleate ice with differing
efficiency. This is commonly known as \emph{site specific} nucleation
behaviour~\cite{vali_interpretation_2014,herbert_representing_2014,vali_technical_2015,vali_repeatability_2008}, and
it can be appreciated to a lesser extent for the other ice nucleating agents considered in
Fig.~\ref{FIG_1}b.  However, it is interesting to note that CHLM crystals seem to lead to two different ice nucleation
regimes, as can be inferred from Fig.~\ref{FIG_1}b (note the two different slopes characterising $n_s$ as a function of
temperature).  As it is becoming increasingly clear that the topography of the ice nucleating agents must play an
important role in the heterogeneous nucleation of ice from liquid
water~\cite{gurganus_nucleation_2014,bi_enhanced_2017,whale_role_2017}, we suspect that structural differences between
the crystalline areas covered by the water droplets are responsible for the wide spread in nucleation temperatures
observed.  This hypothesis is supported by the scanning electron microscope (SEM) images of the (001) face of CHLM
reported in Fig.~\ref{FIG_2}.  While the crystalline plates appear as mostly flat and smooth within the resolution of
$\sim$100 $\mu$m, it is clear that there exist numerous defects, particularly steps and terraces, which can potentially
present opportunities for complicated surface geometries to occur. How exactly the nanometric structure of crystalline
ice nucleating agents affects the kinetics of ice formation is still an open question (see e.g.
Refs.~\citenum{bi_enhanced_2017,campbell_observing_2017}).  In fact, it would be expected that an atomically smooth and
homogeneous CHL surface would nucleate ice with a single nucleation rate and hence within a far narrower range of
temperatures than that reported in Fig.~\ref{FIG_1}.  The role of specific defects and broadly speaking of the surface
morphology to ice formation on CHL - and the vast majority of biological nucleating agents, it thus remains yet to be
fully understood. For instance, it is not immediately clear why CHLM crystals are much more effective than feldspar in
promoting the formation of ice.  In the next section, we will address this issue by showing that in addition to the
topography of the surface, the formation of a peculiar hydrogen bond network at the water-CHLM interface is key in
determining the ice nucleating ability of this compound.

\subsection{\label{r_int} The cholesterol-water interface}

\begin{figure}[htbp]
        \begin{centering}
                \centerline{\includegraphics[width=9.0cm]{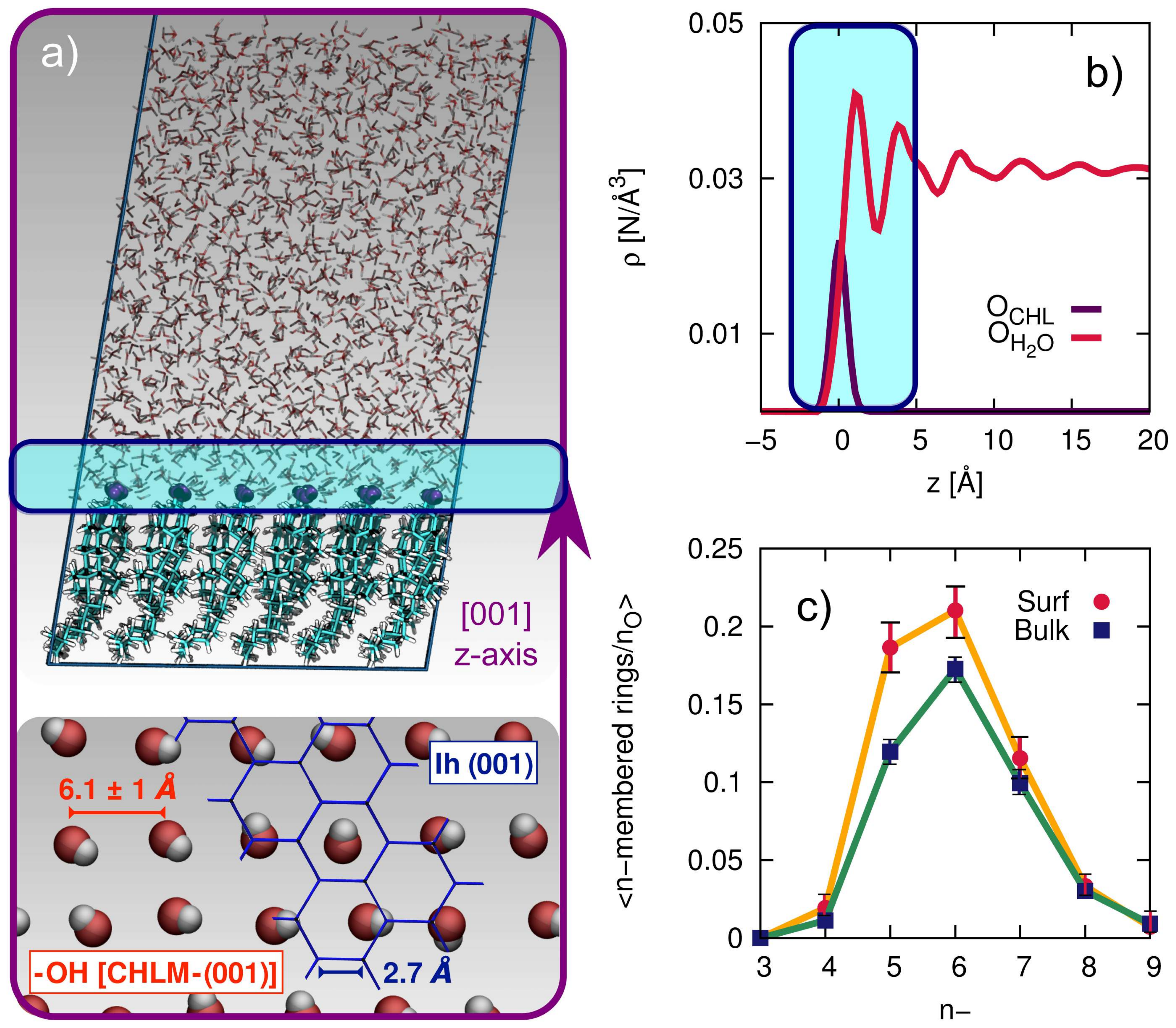}}
        \par\end{centering}
        \protect
        \caption{\textit{Structuring of water on the (001) hydroxylated face of CHLM.} a) Representative snapshot of a
molecular dynamics simulation of a water slab in contact with the (001) hydroxylated face of CHLM. The inset at the
bottom illustrates the arrangement of the hydroxyl group on the CHLM surface; an hypothetical ice $I_h$ (001) plane
(blue) is superimposed on part of the image to highlight the absence of a structural match between -OH groups and ice.
b) Density profile of oxygen atoms belonging to either the -OH hydroxyl groups of CHLM molecules (O$_{CHL}$) or water
molecules (O$_{H_2O})$ along the z-axis parallel to [001] direction, thus normal to the water-CHLM interface. The zero
of the x-axis corresponds to the average position of the $O_{CHL}$ atoms, while the shaded area in green identifies the
water-CHLM interface. Statistics have been accumulated over a 1.5 $\mu$s long simulation at 230 K. c) Number of
n-membered rings of hydrogen bonded water molecules at the water-CHLM interface (Surf) or within the bulk of the water
slab (Bulk), normalised by the number of oxygen atoms in each region. Note that at the water-CHLM interface oxygen atoms
belonging to the CHLM -OH hydroxyl groups have also been considered when computing the rings statistics.}
        \label{FIG_3}
\end{figure}

\noindent In order to investigate the molecular-level details of the CHLM-water interface, we have performed unbiased
molecular dynamics simulations at strong supercooling ($\Delta T_S$ = 42 K) employing the
CHARMM36~\cite{bjelkmar_implementation_2010,lim_update_2012} and the TIP4P/Ice~\cite{abascal_potential_2005} force
fields for CHL and water molecules respectively. Computational details and results concerning the validation of our
computational setup are reported in the Methods section and in the Supplementary Information (SI) respectively, while
the computational geometry is depicted in Fig.~\ref{FIG_3}a.

A water slab ($\sim$ 40 \AA$\ $ thick) is in contact with the hydroxylated (001) surface of CHLM (CHLM$^{-OH}_{001}$),
modelled as a single layer of CHL molecules. This surface is hydrophilic, due to the presence of amphoteric hydroxyl
groups. As CHL molecules are relatively bulky and the crystal is held together by weak electrostatic interactions only,
the arrangement of these -OH groups on the CHLM surface is characterised by a broad distribution of large OH-OH
distances, ranging from 5.1 to 7.1 \AA$\ $ - as illustrated in Fig.~\ref{FIG_3}a. Such a pattern of hydroxyl groups does
not straightforwardly match any particular low-index Miller surface of either hexagonal or cubic ice. This is relevant,
as a good structural match between a substrate and ice~\cite{pedevilla_what_2017} has traditionally been considered as a
"requirement" of an effective ice nucleating agent~\cite{fturn}.

Interestingly, despite the presence of the hydroxyl groups, the density profile of the oxygen atoms of the water
molecules on CHLM reported in Fig.~\ref{FIG_3}b resembles that for water at hydrophobic
walls~\cite{harrach_structure_2014}. The enhancement ($\sim$ 30\% in Fig.~\ref{FIG_3}b) of the density, within the first
peak of the profile, compared to its value in the bulk of the water slab, is much smaller than that (typically a factor
four or six) observed for e.g. water in contact with hydrophilic walls -- or indeed water on kaolinite.  This is
because, the outer layer of the CHLM crystals is much more mobile/flexible than that of kaolinite: this is not
surprising, as we are comparing a molecular organic crystal (held together by van der Waals interactions) with a
(covalently bonded) clay mineral.  Importantly, it is reasonable to assume that a similar degree of flexibility
characterises the majority of organic crystals containing long molecules such as steroids. This is relevant to ice
formation because, as discussed in e.g. Ref.~\citenum{qiu_ice_2017}, the structural fluctuations of organic/biological
ice nucleating particles can strongly affect the kinetics of ice nucleation. In fact, we have shown in
Ref.~\citenum{sosso_ice_2016} that the same argument holds in the case of kaolinite as well: for instance, a ``frozen''
kaolinite surface (atoms are kept fixed during MD simulations) leads to nonphysically fast ice nucleation rates.

\begin{figure}[h!]
        \begin{centering}
                \centerline{\includegraphics[width=8.3cm]{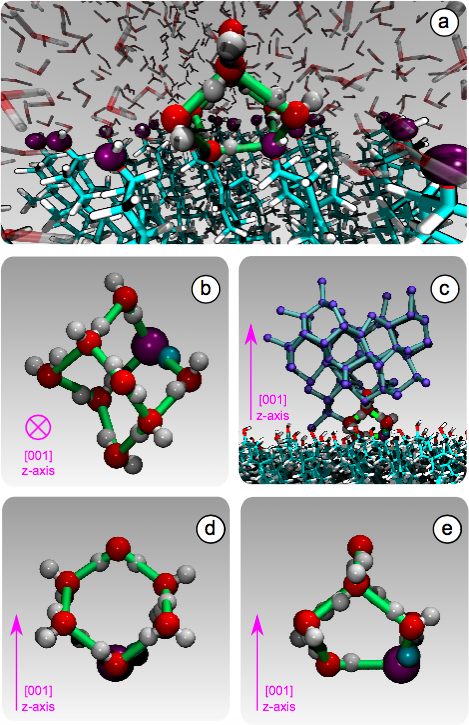}}
        \par\end{centering}
        \protect
        \caption{\textit{The formation of unconventional ice-templating molecular structures: hydrogen bonded {\textit
cages}.} a) A hydroxyl group (purple) of CHLM participates into a hydrogen bonded cage of water molecules.  b) A single
hydrogen bonded cage (top view), which constitutes the building block of cubic ice (panel c, side view).  Note that
these cages are made of both 6-membered (panel d, side view) and 5-membered (panel e, side view) hydrogen bonded rings,
involving water molecules as well as a hydroxyl group provided by the CHLM surface.}
        \label{FIG_4}
\end{figure}

Because of this flexibility of the CHL molecules and the low-density of hydroxyl groups at the water-CHLM interface, we
did not observe (within a 200 ns time scale) the formation of an ordered, ice-like over-layer of water molecules, in
contrast with what is generally found in the case of idealised crystalline surfaces~\cite{fitzner_many_2015},
carbonaceous particles~\cite{lupi_heterogeneous_2014} or kaolinite
crystals~\cite{sosso_ice_2016,sosso_microscopic_2016}.  In fact, most inorganic substrates are characterised by surfaces
where atomic/molecular species are tightly packed, and can thus potentially provide a high density of functional groups
for supercooled water to interact with, typically by forming a more or less ordered overlayer sitting on top of the
crystalline surface. In the case of the water-CHLM interface, however, water molecules can partially infiltrate the
outer layer of the CHLM surface (see Fig.~\ref{FIG_3}a and Fig.~\ref{FIG_3}b) due to the relatively large spacing
between the CHL molecules and the flexibility of the surface itself. As a net result, despite the absence of a flat
overlayer of ice-like water molecules, the amphoteric character of the hydroxyl groups does facilitate the formation of
a network of hydrogen bonded rings of water molecules as well as hydroxyl groups, as illustrated in Fig.~\ref{FIG_3}c.
In particular, we observe the emergence of 6-membered rings of hydrogen bonded water molecules and hydroxyl groups.
These rings are the building blocks of both hexagonal (ice $I_h$) and cubic (ice $I_c$) ice, and are the most abundant
species in bulk water. Note that the occurrence of these rings is actually even more pronounced in the proximity of the
CHLM-water interface (red/orange points/curve in Fig.~\ref{FIG_3}c).  Surprisingly, there is also an increase in the
number of 5-membered water/hydroxyl rings at the crystal-liquid interface. Pentagonal rings are thought to frustrate the
homogeneous formation of ice~\cite{shintani_frustration_2006}; however, in this case both 6- and 5-membered rings alike
contribute to the formation of ice-like fluctuations such as the ''cage'' shown in Fig.~\ref{FIG_4}a. These cages are
indeed the building blocks of ice $I_c$ (see Fig.~\ref{FIG_4}b and Fig.~\ref{FIG_4}c), and involve hydrogen bonds
between water molecules and hydroxyl groups, as depicted in Fig.~\ref{FIG_4}d and Fig.~\ref{FIG_4}e. 
Thus, in this heterogeneous nucleation scenario, the presence of 5-membered rings in not detrimental; on the contrary, they lead to
the formation of ice-like fluctuations of the water network at the water-CHLM$^{-OH}_{001}$ interface.

We note that the emergence of these cages is the reason why we have chosen to consider as the ''interfacial layer''
those water molecules within 5.0 \AA$\ $ from the average position of the oxygens of the CHL hydroxyl groups - as 
illustrated by the shaded region in Fig.~\ref{FIG_3}b. As shown in e.g. Ref.~\citenum{pedevilla_can_2016}, the definition of 
this water layer can have an impact on the analysis of the structure of - in this case - the water-CHLM interface. 
While the interfacial layer can be chosen on the basis on indicators such as the first or second minimum of the 
density profile (see Fig.~\ref{FIG_3}b), we have found that the rather generous cutoff of 5.0 \AA$\ $ is sufficient to accommodate
the substantial extent of the hydrogen bonded cages depicted in Fig.~\ref{FIG_4}d and Fig.~\ref{FIG_3}e. We have also verified that 
by choosing the second minimum of the density profile ($\sim$7\AA) our results, including the trends within the rings statistics
reported in Fig.~\ref{FIG_3}c, are basically unchanged.

Our findings thus contribute the growing body of evidence~\cite{fitzner_many_2015,qiu_ice_2017,bi_enhanced_2017} that the
structural mismatch argument alone cannot be deemed as neither a sufficient nor a necessary criteria to assess, let
alone to predict, the ice nucleating ability of a given substrate~\cite{pedevilla_what_2017}. This is bound to be
especially true in the case of biological ice nucleating agents such as macromolecules~\cite{pummer_suspendable_2012},
where the notion itself of a lattice mismatch is ill defined. In fact, we argue that organic crystals such as
cholesterol lie halfway in between inorganic (e.g. mineral crystals) and biological (e.g. bacterial fragments) ice
nucleating agents, as they are characterised by the relatively flat and (in this case) -OH regularly patterned surfaces
of the former while showing the flexibility of the latter. This is particularly relevant for CHL, which is a substantial
component of animal cellular membranes~\cite{krause_structural_2014} and could thus contribute to promote the
heterogeneous formation of ice in biological matter -- a possibility we will investigate in future work.  In this
respect, it is interesting to note that very recent simulations~\cite{hudait_ice-nucleating_2018} suggest that ice can
bind to antifreeze proteins via ``anchored clathrate'' motifs not dissimilar to the molecular cages discussed above.

\subsection{\label{r_ffs} Ice nucleation mechanism and kinetics}

\noindent In order to characterise the mechanism as well as the kinetics of ice nucleation on CHLM$^{-OH}_{001}$ we have
performed forward flux sampling (FFS)
simulations~\cite{allen_forward_2009,li_homogeneous_2011,li_ice_2013,valeriani_rate_2005,wang_homogeneous_2009,bi_probing_2014,gianetti_computational_2016,haji-akbari_computational_2017}.
While other enhanced sampling techniques are in principle available, such as metadynamics~\cite{laio2002escaping},
transition path sampling~\cite{Bolhuis_AnnuRevPhysChem_2002_TPS}, and seeded molecular
dynamics~\cite{Espinosa_JChemPhys_2016_Homo-Seeding}, FFS represents a ``gold standard'' approach when dealing with ice
nucleation (see e.g.
Refs.~\citenum{haji-akbari_direct_2015,bi_heterogeneous_2016,sosso_microscopic_2016,haji-akbari_computational_2017,haji-akbari_forward-flux_2018}).
This method involves partitioning the path from (in this case) liquid water to ice, described by an order parameter
$\lambda$, into a collection of interfaces $\lambda_i$. Here, $\lambda$ corresponds to the number of water molecules
within the largest ice nucleus, which can be located either in the bulk of the water slab or at the
water-CHLM$^{-OH}_{001}$ interface. A diffuse crystal-liquid interface has been taken into account into the definition
of $\lambda$, which relies on local bond order parameters (see SI and Ref.~\citenum{tribello_analyzing_2017}),
consistent with Ref.~\citenum{sosso_microscopic_2016}.  Starting from the natural fluctuations of liquid water toward
the ice phase, i.e. pre-critical ice nuclei as sampled within $\mu$s long unbiased MD simulations, the nucleation rate
$\mathcal{J}$ can be obtained as the product of the flux $\Phi_{\lambda_0}$ by which the system reached the first
interface $\lambda_0$, times the product of the sequence of the individual crossing probabilities
$P(\lambda_i|\lambda_{i-1})$:

\begin{equation}
        \mathcal{J}={\Phi}_{\lambda_0}\prod_{i=1}^{N_{\lambda}} P (\lambda_i|\lambda_{i-1})
\end{equation}

\noindent In this way, the (exceedingly small) total probability $P(\lambda_{ice}|\lambda_{0})$ for a certain MD
trajectory to reach the ice basin is decomposed in a collection of (manageable) crossing probabilities which we compute
by a large number (10$^3$ to 10$^5$) of unbiased MD trial runs from $\lambda_{i-1}$ to $\lambda_i$. Further details
about the FFS algorithm can be found in the SI. We note that we have used the same water model (TIP4P/Ice) at the same
strong supercooling ($\Delta T_S$=42 K) as employed previously to compute the homogeneous ice nucleation rate and the
heterogeneous ice nucleation rate on kaolinite, a clay mineral of relevance to atmospheric science. As such, we can
compare directly our results with those of Refs.~\citenum{haji-akbari_direct_2015} and
~\citenum{sosso_microscopic_2016}.

From the very early stages of the nucleation process, we observe a strong preference for ice to form at the
water-CHLM$^{-OH}_{001}$ interface -- as opposed to within the bulk of the water slab. In fact, $\sim$ 75\% of the
pre-critical ice nuclei we observe as natural fluctuations of the supercooled water network ($\lambda$=0) sit on top of
the CHLM$^{-OH}_{001}$ surface. The calculated growth probability $P(\lambda|\lambda_{0})$ as a function of lambda,
together with the fraction of ice nuclei that can be found at the water-CHLM$^{-OH}_{001}$ interface are reported in the
SI (Fig. S2b). By the time the FFS algorithm has reached $\lambda$=125, no nuclei within the bulk of the water slab
survive.  We have observed a similar trend in the case of ice nucleation on the hydroxylated (001) basal face of
kaolinite~\cite{sosso_microscopic_2016}, but the fraction of ice nuclei at the water-kaolinite interface at the initial
stages of the FFS algorithm was much smaller ($\sim$ 25\%). This suggests that pre-critical ice-like fluctuations, which
we have recently investigated in the broader context of heterogeneous crystal
nucleation~\cite{fitzner_pre-critical_2017}, are much more likely to occur at the surface of CHLM compared to kaolinite.

\begin{figure}[t!]
        \begin{centering}
                \centerline{\includegraphics[width=8.3cm]{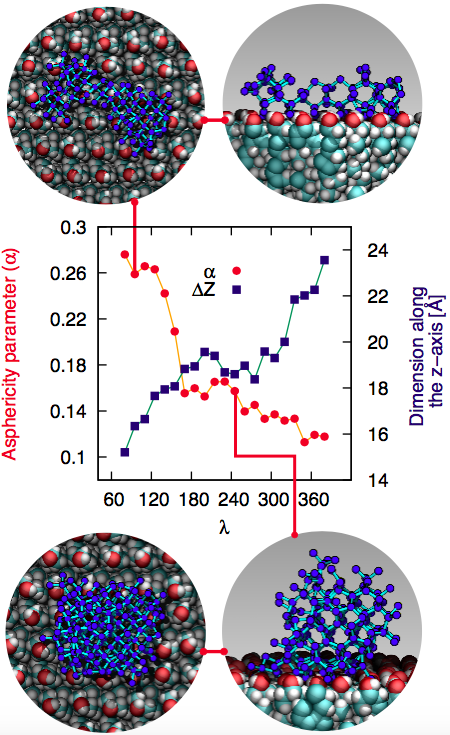}}
        \par\end{centering}
        \protect
	\caption{\textit{The early stages of ice nucleation at the water-CHLM$^{-OH}_{001}$ interface involve
non-spherical ice crystallites.} Asphericity parameter $\alpha$ and spatial extent of the ice nuclei along the direction
normal to the CHLM slab $\Delta Z$ as a function of $\lambda$ for ice nuclei at the water-CHLM$^{-OH}_{001}$ interface.
The insets correspond to top and side views of typical ice nuclei forming at the water-CHLM$^{-OH}_{001}$ interface,
containing about 100 (top) and 245 (bottom) water molecules.}
        \label{FIG_5}
\end{figure}

The mechanism of ice nucleation  at the water-CHLM$^{-OH}_{001}$ interface is illustrated in Fig.~\ref{FIG_5}: the early
stages involve the formation of elongated, almost one dimensional, linear, chain-like ice nuclei preferentially along
specific directions (see SI), due to the particular arrangement of the -OH hydroxyl groups on the CHLM$^{-OH}_{001}$
surface. However, larger nuclei (corresponding to increasing values of $\lambda$) progressively assume a more isotropic
shape, as indicated by the evolution of the asphericity parameter $\alpha$ (equal to 1 and 0 for a infinitely elongated
rod and a perfect sphere respectively) as a function of $\lambda$. At the same time, the 1D character of the nuclei
evolves toward a more compact geometry, with a significant growth along the [001] direction (z-axis) normal to the
water-CHLM$^{-OH}_{001}$ interface, as demonstrated by the trend of the dimension $\Delta Z$ of the ice nuclei along
that axis, also reported in Fig.~\ref{FIG_5}. The resulting morphology of the ice crystals, though, remains to be
investigated because of the emergence of finite size effects. Overall, the evolution of the ice nuclei within the early
stage of ice nucleation at the water-CHLM$^{-OH}_{001}$ interface possesses some similarities with the case of ice
formation on kaolinite~\cite{sosso_microscopic_2016}, where ice nuclei spread into a 2D, planar geometry before stacking
additional ice layers along the normal to the water-kaolinite interface, once the critical nucleus has been reached.
Thus, these findings suggest that the nature of the early stages of heterogeneous ice nucleation at strong supercooling
($\Delta T_S$ = 42 K) has a strong anisotropic character, in stark contrast with the assumptions prescribed by classical
nucleation theory (CNT)~\cite{cabriolu_ice_2015}. 

In fact, CNT does not take into account the molecular structure nor the ''chemistry'' of the substrate: these aspects
are only implicitly included into the value of the contact angle of the ice nuclei with respect to the substrate.
However, microscopic features such as the particular arrangement of the hydroxyl groups on the CHLM surface can
influence the shape and the energetics of the ice nuclei. In the case of ice on CHLM, water molecules at the
water-cholesterol interface find convenient to harness the directionality of the -OH pattern (see Fig.~\ref{FIG_3}a and
Fig.~\ref{FIG_3} in the SI) to form anisotropic ice nuclei (see Fig.~\ref{FIG_5} and Fig.~\ref{FIG_3} in the SI), which
are likely to be characterised by a much smaller interfacial energy if compared to the hemispherical shape predicated by CNT in the
case of perfectly flat, featureless substrates. We note however that in order to probe this aspect of CNT
quantitatively, it would be desirable to improve the current enhanced sampling techniques to take into account milder
supercooling - and thus larger critical ice nuclei.

The ice nucleation rate on the CHLM$^{-OH}_{001}$ surface obtained from our FFS simulations is 10$^{27\pm 3}$
s$^{-1}$m$^{-3}$, about 20 orders of magnitude larger than the homogeneous ice nucleation rate at the same supercooling
-- calculated via FFS simulations using the same water model~\cite{haji-akbari_direct_2015}.  This spectacular
enhancement of the kinetics of ice formation is due to the small heterogeneous critical nucleus size $N^*_{\text{H}}$,
which we estimate (as discussed in detail in the SI) to contain 250 $\pm$ 50 water molecules -- a number consistent with
the predictions of CNT (see SI and Ref.~\citenum{sosso_microscopic_2016}). Interestingly, these results are very similar
to what we have previously obtained in the case of ice formation on kaolinite~\cite{sosso_microscopic_2016}, where we
calculated $\mathcal{J}$ = 10$^{26\pm 2}$ s$^{-1}$m$^{-3}$ and $N^*_{\text{H}}$ = 225 $\pm$ 25.  However, it has to be
said that the FFS simulations performed in this work (as opposed to the case of kaolinite~\cite{sosso_microscopic_2016})
may be suffering from finite size effects (discussed in the SI), which could both enhance the kinetics of ice nucleation
(as the ice nuclei feel the influence of their periodic images) and/or hamper the growth of ice crystals (as the
simulation box most likely does not match the periodicity of the growing ice crystal). Our estimates of $\mathcal{J}$
and $N^*_{\text{H}}$ have therefore to be taken with care.

The fact that the kinetics of ice formation on the CHLM$^{-OH}_{001}$ surface seems to be comparable with that of an
inorganic crystal such as kaolinite is not entirely unexpected, as the (001) hydroxylated surface of kaolinite also
presents -OH groups at the water-crystal interface which are capable of templating the formation of ice-like structures.
However, supercooled water on kaolinite forms a dense, hexagonal ordered overlayer of ice-like molecules sitting on top
of the hydroxyl groups~\cite{sosso_ice_2016}, while, as we have discussed in the previous section, water molecules can
partially infiltrate the CHLM$^{-OH}_{001}$ surface to form 5- and 6- membered hydrogen bonded rings, resulting in a
much less ordered and way less dense overlayer. As both substrates (kaolinite and CHLM) are characterised by the
presence of hydroxyl groups which facilitate the formation of ice, the much faster kinetics of ice nucleation we have
observed experimentally for CHLM compared to kaolinite (especially at mild supercooling, Fig.~\ref{FIG_1}b) is likely to
be due to the different surface topography of the two compounds.

\subsection{\label{comp} Competition between cubic and hexagonal ice}

\begin{figure*}[htbp]
        \begin{centering}
                \centerline{\includegraphics[width=18.75cm]{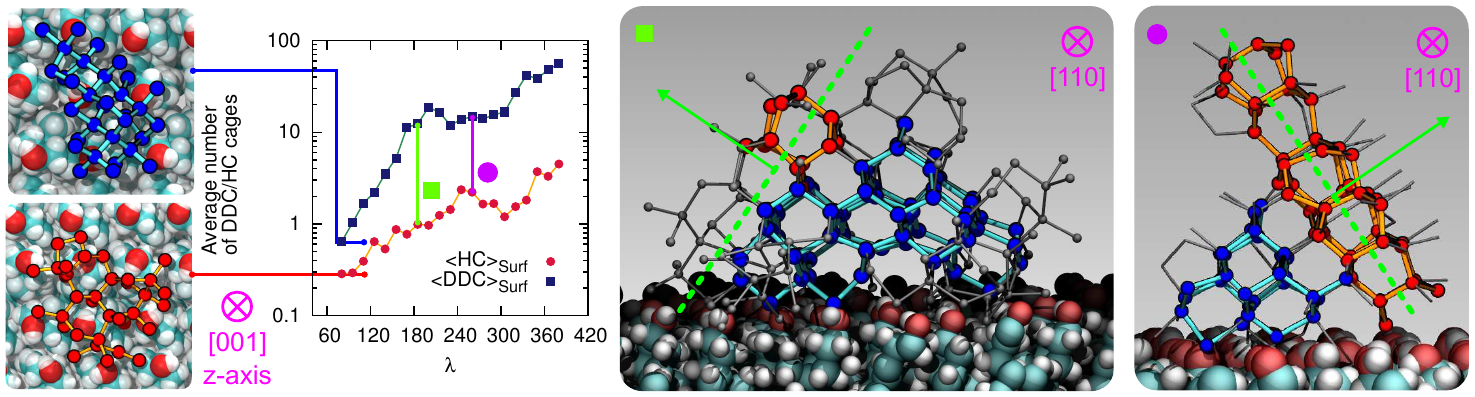}}
        \par\end{centering}
        \protect
	\caption{\textit{Competition between cubic (blue/cyan spheres/sticks) and hexagonal (red/orange spheres/sticks)
ice within the early stage of ice nucleation at the water-CHLM$^{-OH}_{001}$ interface}. The average number of double
diamond and hexagonal cages (DDC and HC, the building block of ice $I_c$ and ice $I_h$ respectively) is reported as a
function of the order parameter $\lambda$. Insets on the left show representative ice $I_c$ and ice $I_h$ fluctuations
(top view) at the first FFS interface ($\lambda$=80).  Insets on the right show representative ice nuclei at
$\lambda$=165 and 260, where the competition between the two polymorphs becomes more evident. The dashed (green)
lines/arrows indicate the crystallographic plane/direction along with ice $I_h$ has the possibility to grow on top of
ice $I_c$.}
        \label{FIG_6}
\end{figure*}

\noindent At the strong supercooling considered here ($\Delta T_S$ = 42 K), homogeneous ice nucleation results in a
mixture of ice $I_c$ and ice $I_h$ known as stacking disordered ice
$I_{sd}$~\cite{moore_is_2011,hansen_formation_2008,malkin_stacking_2014}.  However, things can be quite different in the
heterogeneous case. For instance, the hydroxylated (001) basal face of kaolinite promotes exclusively the formation of
the primary prism face of ice $I_h$~\cite{sosso_microscopic_2016,sosso_ice_2016}.  In the case of the CHLM$^{-OH}_{001}$
surface, we observe both ice $I_c$ and ice $I_h$ nuclei at the very early stages of the nucleation process, as depicted
in the inset (left side) of Fig.~\ref{FIG_6}.  These ice-like fluctuations originate from the templating effect of the
hydroxyl groups on the CHLM surface, as illustrated in Fig.~\ref{FIG_4}a-e (and Fig.S3 in the SI).  In principle, even
if ice $I_c$ nuclei are three times more abundant than ice $I_h$ nuclei at the first FFS interface ($\lambda=80$), we
would expect the formation of ice on the CHLM$^{-OH}_{001}$ surface to proceed via the growth of ice $I_{sd}$. In fact,
as shown in Fig.~\ref{FIG_6}, the competition between the growth of two ice polytypes at 230 K (i.e. $\Delta T$=42 K) is
dominated by ice $I_c$: by the time the ice nuclei have reached a post-critical size (e.g.  $\lambda=360$), the average
number of Hexagonal Cages (HC~\cite{haji-akbari_direct_2015}, the building blocks of ice $I_h$) is still about three
times larger than that of DDC (Double Diamond Cages~\cite{haji-akbari_direct_2015}, the building blocks of ice $I_c$).

Interestingly, despite the predominance of ice $I_c$ within the growing ice nuclei, ice $I_h$ can still form and grow
along a specific direction (the [111] of the cubic phase) on top of ice $I_c$ crystals (which in turn grow along the
[100] direction, normal to the plane of the water-CHLM$^{-OH}_{001}$ interface), as illustrated in the insets (right
side) of Fig.~\ref{FIG_6}. The coexistence of ice $I_c$ and ice $I_h$ is thus likely to result in ice $I_{sd}$ crystals
at strong supercooling. However, at milder supercooling ice $I_h$ fluctuations are expected to become more relevant, and
in fact experimental evidence suggests that the macroscopic crystalline habit of ice crystals grown on CHLM at $\Delta
T$=2 K is indeed that of ice $I_h$~\cite{fukuta_experimental_1966}. Importantly, we did not observe such a competition
between ice $I_c$ and ice $I_h$ in the case of kaolinite, where the cubic polytype is basically absent throughout the
whole nucleation process~\cite{sosso_microscopic_2016,sosso_ice_2016}. In fact, we argue that, in the case of CHLM
crystals, different nucleation sites (whose exact nature remains to be determined) could promote chiefly ice $I_c$ or
ice $I_h$ according to the different degree of supercooling, thus contributing to unravel the strong ice nucleating
ability of CHLM crystals along such a wide range of temperatures. This argument would suggest that the multi-component
nature of ice nucleation on biological matter could be at least partially attributed to a greater variety of nucleation
sites -- as well as the specific templating effect of functional groups acting as hydrogen bond donor and/or acceptors
with respect to supercooled liquid water. Moreover, we have shown in this work that some of these functional groups -
such as the hydroxyl groups characterising the water-CHLM$^{-OH}_{001}$ interface - can even promote a different ice
polytype at the same time, possibly according to different supercooling.

Finally, we note that, in agreement with previous simulations of ice nucleation~\cite{sosso_ice_2016,qiu_ice_2017}, the
flexibility of the CHLM$^{-OH}_{001}$ has an impact on the extent and the structure of the ice-like fluctuations at the
CHLM$^{-OH}_{001}$-water interface, and that the anhydrous crystalline phase of CHL also displays substantial ice
nucleating potential. These two aspects are both addressed in detail in the SI.

\section{\label{disc} Conclusions}

\noindent By means of a blend of experiments and simulations, we have unravelled the the origins of ice nucleation on
cholesterol (CHL), a prototypical organic crystal of relevance to cryopreservation.  Our results suggest that its
exceptional ice nucleating activity stems from the ability of its flexible hydrophilic surface to form unconventional
ice-templating structures -- specifically, hydrogen bonded cages comprising 6- as well as 5-membered rings.  In
addition, the experimental evidence reported here suggests that the intrinsic potential of cholesterol to nucleate ice
may potentially be enhanced by specific topological features of the crystalline habit.  In particular, droplet freezing
measurements show that cholesterol promotes the heterogeneous formation of ice across a wide range of temperatures (from
-4 to -20 C\degree).  In fact, we find that CHLM crystals nucleate ice far better than the mineral feldspar, which is
one of the most effective inorganic ice nucleating agents of relevance to atmospheric science.  Moreover, electron
microscopy measurements suggest that the broad range of freezing temperatures we observe for CHLM crystals may be due to
the coexistence of diverse structural features of the crystalline surface, which in turn can act as different nucleation
sites. The microscopic structure of the latter remains to be assessed, but the possibility that different parts of the
CHLM surface may nucleate ice with different efficiency suggests that surface topography can play an important role in
determining the ice nucleating ability of organic crystals.

Surprisingly, we find that CHLM crystals, despite being exceptionally good ice nucleating agents, do not provide a
conventional template for ice to form. Specifically, molecular simulations reveal that, as opposed to what has been
reported for supercooled water in contact with simple model substrates (e.g. Lennard-Jones crystals, which allow
to rapidly explore different surface geometries~\cite{fitzner_many_2015}) and/or inorganic materials
(such as carbonaceous particles~\cite{lupi_heterogeneous_2014}, or clay
minerals~\cite{sosso_ice_2016,sosso_microscopic_2016}), water on the (001) hydroxylated surface of cholesterol
monohydrate (the most abundant interface in aqueous environments) does not form an ordered, dense, ice-like overlayer.
Instead, due to the flexibility of the CHLM surface and its relatively low density of hydroxyl groups, water molecules
partially infiltrate the crystal, forming a network of both 6- and 5-membered hydrogen bonded rings. The latter involve
water molecules as well as hydroxyl groups provided by CHL molecules. While some of these structural features
(particularly pentagonal rings) are known to hinder homogeneous water freezing, we find that they actually facilitate
the heterogeneous formation of both hexagonal and cubic ice on CHLM crystals. In fact, enhanced sampling simulations
suggest the emergence of stacking disordered ice (a mixture of the two polytypes) at the water-CHLM interface.  This is
in stark contrast with what we have previously observed in the case of e.g. the clay mineral kaolinite, where only the
hexagonal polytype was observed along the whole nucleation process~\cite{sosso_microscopic_2016}. In fact, more often
than not a given crystalline substrate nucleates exclusively one of the two ice
polytypes~\cite{sosso_crystal_2016,kiselev_active_2017}.  Moreover, we find that the nucleation rate of ice on CHLM
crystals is basically identical to that we have previously calculated in the case of kaolinite - at the same strong
supercooling ($\Delta T_S$ = 42 K). Kaolinite and CHLM are both characterised by an hydrogen bond network capable of
facilitating the formation of ice nuclei: thus, the substantial difference in the ice nucleating ability we observe
experimentally for these two compounds is most likely rooted into their surface topography.
In fact, the $n_s$ data reported in Fig.~\ref{FIG_1}b suggest that two populations of potentially different ice nucleating sites may
coexist on the CHLM surface. The change in the slope of the CHLM data is reminiscent of that observed for freezing
spectra for birch pollen~\cite{osullivan_relevance_2015,pummer_suspendable_2012}, 
which has been attributed to the presence of two different ice nucleating
macromolecules~\cite{augustin_immersion_2013}. Similarly, we argue that there may be two
different broad classes of ice nucleating sites on CHLM, represented by the two different slopes in the freezing
spectra. Due to the spatially sporadic nature of the highly active sites, which are not present in every millimetre
diameter droplet, it seems likely that these two different classes of ice nucleation sites are related to specific
defects or the diverse topography of the CHLM, rather than any factors related to the bulk molecular structure of CHLM.

In addition, the emergence of stacking disordered ice phases during the heterogeneous formation of ice has been experimentally
observed~\cite{malkin_stacking_2014}, and consequently ascribed to different crystal growth regimes. Our results offer
the intriguing prospect that the nucleation process itself may favour, in some cases, the formation of stacking
disordered ices. Thus, we argue that the dramatic ice nucleation ability of certain organic materials may be traced down
not only to the formation of a network of hydrogen bonds between water and the nucleation sites, but also to the
capability of specific surfaces to promote at the same time different ice polytypes as a function of supercooling. In
order to verify this hypothesis, though, we would need to investigate ice nucleation on CHLM at milder supercooling. To
this end, an heterogeneous seeded molecular dynamics approach is currently being validated~\cite{HSEED}.  Our results
also suggest that organic crystals sit in between inorganic and biological materials, when it comes to promoting the
formation of ice: substrates like CHLM are characterised by relatively flat surfaces exposing an array of amphoteric
functional groups, much like several inorganic ice nucleating agents (e.g. kaolinite, feldspar, hydroxylated graphene),
but the flexibility of the surface and the low density of such functional groups is typical of biological nucleating
agents such as macromolecules and bacterial fragments.  This is especially relevant in the case of CHL, a molecule which
is not only used in crystalline form as an ice nucleating agent in cryopreservation applications, but that significantly
contributes to the composition of animal cell membranes as well.

In summary, the experiments and simulations presented in this work indicate that cholesterol crystals are incredibly
efficient ice nucleating agents, active across a broad range of supercooling.  We show that such strong ice nucleating
activity is due to the intrinsic potential of the flexible amphoteric surfaces of CHLM to form unconventional
ice-templating molecular structures. It is likely that microscopic structural features of the crystals could further
enhance the ability of CHLM (and potentially of other organic crystals) to form ice, by offering a diverse array of
nucleating sites. In fact, we believe that for an ice nucleating agent to be very efficient, a combination of
interfacial ``chemistry'' and surface topography is generally required.  This interplay could thus be the key to
understand the heterogeneous formation of ice on molecular organic crystals, and it may provide a starting point for the
investigation of ice in soft and biological matter at the molecular level.  In particular, tailoring the microscopic
structure of the substrate and modifying the nature as well as the density of hydrogen-bonding functional groups at the
water-substrate interface can be seen as two different routes to engineer the ice nucleating ability of novel
cryoprotectants, the design of which, at the moment, largely relies on the high-throughput screening of whole libraries
of different compounds. The absence of a proper structure-to-function paradigm is perhaps the most pressing challenge in
cryobiology: this is why future work will be devoted to assess whether and how hydrogen-bonding functional groups other
than hydroxyls would be equally effective to enhance the kinetics of heterogeneous ice nucleation.

\section*{Conflicts of interest}
There are no conflicts to declare.

\section*{Acknowledgements}
This work was supported by the European Research Council (ERC) under the European Union's Seventh Framework Programme:
Grant Agreement number 616121 (HeteroIce project [AM, PP, GCS]), Grant Agreement number 632272 (IceControl project [TFW,
BJM]) and Grant Agreement number 648661 (MarineIce project [TFW, BJM]).  MAH and BJM are also supported by the
Engineering Physical Sciences Research Council (EPSRC) via the Research Grant number EP/M003027/1.  GCS and PP
acknowledge the use of the UCL Grace and Legion High Performance Computing Facilities, the use of Emerald, a
GPU-accelerated High Performance Computer, made available by the Science \& Engineering South Consortium operated in
partnership with the STFC Rutherford-Appleton Laboratory and the use of ARCHER UK National Supercomputing Service
(http://www.archer.ac.uk) through the Materials Chemistry Consortium via the EPSRC grant number EP/L000202.
The Forward Flux Sampling simulations were supported by a grant from the Swiss National Supercomputing Centre (CSCS; project ID s758). 

\section*{Supplementary Information}
\noindent Supplementary Information (SI) can be found at: {\tiny http://www.rsc.org/suppdata/c8/sc/c8sc02753f/c8sc02753f1.pdf} 

%
%
%


\end{document}